# System-Self as a Data Structure: An Architectural Approach to Bounded Adaptation

Erwin Franz*, Alhassan Yasin

*The Johns Hopkins Applied Physics Laboratory, Laurel, MD 20723, United States*
*The Johns Hopkins University, Baltimore, MD 21218, United States*

**Abstract**

Safety critical autonomous systems often adapt by adjusting controller parameters while keeping the underlying architecture fixed. This strategy breaks down when shifts in sensing, resource availability, or component health invalidate the original structural assumptions. This work introduces a method in which systems maintain an explicit, graph-based representation of their architecture and reason over it during operation. The system is modeled as a directed graph of physical, functional, and model based modules, with edges capturing information and control dependencies. Adaptation is posed as a joint optimization over architectural configurations and module parameters, subject to operational constraints using a Monitor–Analyze–Plan–Execute loop. Performance degradation is isolated via residual decomposition and dependency weighted influence propagation, and candidate adaptations are filtered using a stability aware mechanism. The approach is demonstrated on a differential drive robot under sensor drift and actuator faults. A fixed architecture accumulates tracking errors of up to 24 m and 13 m, respectively, whereas architecture aware adaptation reduces error to under 1.5 m in each case, by selecting fault appropriate configurations. These results show the value of reasoning over system structure, while preserving stability, rather than relying solely on parameter tuning.

## 1. Introduction

Existing adaptive techniques primarily address uncertainty through parameter adaptation. In classical adaptive control, controller parameters are adjusted online to compensate for modeling errors or disturbances while maintaining a fixed controller structure [1, 2]. Similarly, learning-based approaches such as reinforcement learning and neural network controllers improve performance by adapting parameters within a predefined computational architecture [3]. However, changes in sensing availability, resource constraints, or task requirements may render the original system structure suboptimal or infeasible, requiring modifications to the architecture itself rather than parameter adjustment alone. Biological systems demonstrate a broader form of robustness. Inspired by this, Bongard et al. showed that robots can recover from damage by constructing internal models of their own morphology through self-directed exploration [4], highlighting the importance of internal self-representation for structural adaptation.

* Corresponding author.
*E-mail address:* erwin.franz@jhuapl.edu

Motivated by these ideas, we consider an approach in which autonomous systems maintain an explicit internal representation of their architecture as a structured graph. Nodes represent physical components, functional processes, and system models; edges represent dependencies governing information and control flow. By representing system structure as a runtime data structure, the architecture becomes an object that can be evaluated, modified, and selected according to system objectives and operational constraints. Adaptive behavior is then expressed as a joint optimization over architectural configurations and module parameters, enabling the system to reason over how it should be organized, not only how its parameters should be tuned.

The remainder of this paper introduces the architectural representation, formulates the joint adaptation problem, and demonstrates the approach on a differential drive robot performing trajectory tracking under changing operating conditions.

## 2. Related Work

Adaptive control methods such as Model Reference Adaptive Control adjust controller parameters online while preserving a fixed structure [1, 2]. Learning-based approaches similarly optimize parameters within a predefined architecture [3], and fault-tolerant control extends this by enabling reconfiguration through predefined switching logic [10]. Across these approaches, adaptation is confined to a pre-set collection of control structures rather than supported by explicit reasoning over architectural representations. Self-modeling robotics offers a closer conceptual connection. Bongard et al. showed that robots can infer their own morphology through self-directed experimentation to compensate for physical damage [4]. In contrast, the present work encodes subsystem properties directly as attributes within a graph-based architectural model, enabling structural reasoning without exploratory identification. Self-adaptive software frameworks such as Rainbow [11] and autonomic computing architectures [12] apply the Monitor–Analyze–Plan–Execute over a Knowledge base (MAPE-K) loop pattern to modify system structure at runtime, treating architecture as a reconfigurable artifact. Similar principles have been explored in robotic systems using explicit runtime models and metacontrol for architectural reconfiguration [13], typically relying on discrete model-based reasoning over predefined configurations. Architectural representations in systems engineering, such as Design Structure Matrices [5, 6], capture component dependencies but are typically used at design time. The representation introduced here is intended for runtime use and incorporates stability-aware action filtering grounded in Lyapunov stability theory [7] and eigenvalue-based margin conditions for linear systems [1], ensuring that architectural transitions preserve closed-loop stability. Meta-learning [9] and active inference [8] approaches are complementary but focus on parameter-level adaptation within fixed computational structures rather than reasoning over system organization.

## 3. Problem Formulation

### *3.1 System Architecture and Adaptation Model*

Traditional adaptive control and learning-based approaches adjust system parameters while the underlying architecture remains fixed. Let $\mathbf{a}$ denote the system architecture and $\boldsymbol{\theta}$ the set of parameters associated with that architecture. Let $\mathbf{e} \in \mathbf{E}$ represent environmental conditions that influence system performance. The objective of the control or decision mechanism is to maximize a performance function $\mathbf{J}(\cdot)$ that captures task-level objectives such as tracking performance, stability, or mission success. Under the conventional formulation, adaptation occurs only within the parameter space associated with a fixed architecture. This can be written as

$$\theta^* = argmax_{\{\theta \in \Theta_a\}} J(a, \theta, e) \quad (1)$$

where $\boldsymbol{\Theta_a}$ denotes the admissible parameter space for architecture **a**. In this formulation, the architecture remains constant while the system adapts by modifying controller gains, policy parameters, or model coefficients.

When sensing availability, resource constraints, or system degradation render the current architecture suboptimal, parameter adaptation alone is insufficient. We therefore consider a broader formulation in which both architecture and parameters may vary.

Let **A** denote the set of admissible architectures. The adaptive design problem can then be expressed as

$$(a^*, \theta^*) = argmax_{\{a \in A, \theta \in \Theta_a\}} J(a, \theta, e) \quad (2)$$

subject to system constraints

$$C(a, \theta, e) \leq 0 \quad (3)$$

where **C(·)** represents safety, resource, or operational constraints. These constraints may capture limits on computation, energy consumption, latency, or safety-critical conditions. In this formulation, adaptation involves reasoning over both the architecture space **A** and the parameter space associated with each architecture.

Enabling architectural reasoning requires an explicit representation of system structure. We represent a system architecture as a structured composition of modules and their interconnections

$$a = (M, \Gamma) \quad (4)$$

where **M** = $\{m_1, m_2, \ldots, m_n\}$ denotes the set of system modules and **Γ** represents the interconnection structure among those modules. Modules may correspond to sensing components, estimation algorithms, controllers, planners, or other functional subsystems. Each module encodes its functional type, parameter space, resource requirements, and operational constraints, and exposes its capabilities through defined interfaces. Combined with environmental conditions and system policies, this representation defines a bounded architectural configuration space that enables the system to evaluate alternative module configurations and select among admissible architectures satisfying performance objectives and constraints.

### *3.2 System Architecture Representation*

To enable reasoning over architectural alternatives, the system architecture must be represented explicitly. The architecture is modeled as a directed graph that captures both structural and functional elements within a single unified representation. Physical components, functional processes, and system models are represented as modules of the same graph, enabling the system to reason about functional organization consistently with the underlying system configuration. Let the set of modules composing the system be defined as

$$M = \{m_1, m_2, \ldots, m_n\} \quad (5)$$

where each module $m_i$ represents either a physical subsystem or a functional process within the system. Each module is characterized by attributes that describe its role and operational properties. A module can be defined as:

$$m_i = (\tau_i, \Theta_i, k_i) \quad (6)$$

where $\tau_i$ denotes the functional type of module $m_i$, $\Theta_i$ denotes the parameter space associated with the module, and $k_i$ denotes operational or compatibility constraints. Modules may represent physical subsystems, functional processes, or system models used by those processes.

$$M = M_P \cup M_F \cup M_M \quad (7)$$

where $M_P$ denotes the set of physical modules, $M_F$ denotes the set of functional modules, and $M_M$ denotes modeling nodes. Physical modules correspond to tangible components such as sensors, actuators, and structural elements of the system. Functional modules represent processes such as estimation, control, or planning. Modeling nodes represent internal models used for prediction, estimation, or adaptation.

The interconnection structure Γ defines the dependencies and information flow between modules

$$\Gamma \subseteq M \times M \tag{8}$$

An ordered pair $(m_i, m_j) \in \Gamma$ indicates that module $m_i$ provides information, influence, or functionality required by module $m_j$. These directed connections define the flow of sensing data, state estimates, control commands, and task-level information within the system.

To support computational reasoning over the architecture, dependencies between modules may include additional attributes describing the relationship between connected modules. These attributes can capture properties such as communication latency, uncertainty, resource usage, or dependency strength. The architectural representation can therefore be interpreted as an attributed directed graph

$$G = (M, \Gamma) \tag{9}$$

where nodes are system modules and edges are directed dependencies. Edge and node attributes capture interaction properties — such as dependency strength or resource usage — enabling the system to prioritize modules during adaptation based on their influence on performance. Given an architecture defined by its modules and interconnections, the parameter space associated with that architecture can be expressed as

$$\Theta_a = \Theta_1 \times \Theta_2 \times \ldots \times \Theta_n \tag{10}$$

where $\Theta_i$ denotes the parameter space associated with module $m_i$. The parameter vector θ used in the adaptive formulation therefore consists of the parameters associated with the modules composing the architecture. Fig. 1 illustrates this expanded graph representation, providing a unified description of system structure, functional organization, and internal models of a differential drive robot.

## 4. Architecture-Aware Adaptation Experiment

To illustrate the proposed architectural representation and adaptation approach, a differential drive robot is evaluated in a trajectory tracking task. The robot follows a reference trajectory generated using spline interpolation, producing a smooth path between an initial pose and a goal location. The objective of the experiment is to evaluate whether the proposed representation enables the system to adjust its architecture during execution when operating conditions change.

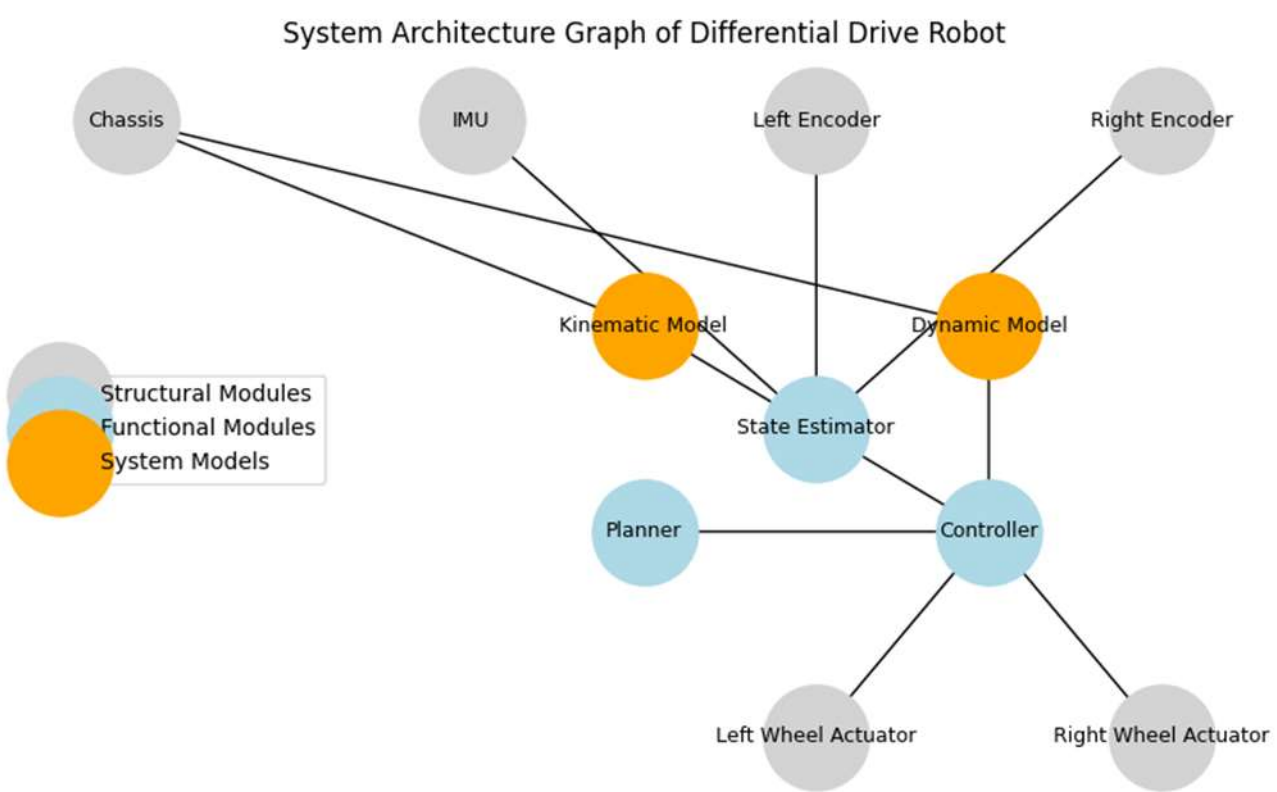

**Figure 1.** Form–function architecture graph of a differential drive robot. Nodes represent system modules $m_i$ (structural, functional, and model components); directed edges represent architectural dependencies $(m_i, m_j) \in \Gamma$ defining information and control flow.

*4.1 Architectural Representation of the Robot*

The robot architecture follows the representation in Eq. (4), comprising three module types: structural (chassis, actuators, sensors), functional (estimation, planning, control), and model (kinematic and dynamic models). Directed edges encode information flow including sensor data, state estimates, and control commands, as illustrated in Fig. 1. This representation enables the system to analyze module dependencies and reason over alternative configurations when performance degrades.

*4.2 Architecture-Aware Reasoning Process*

The system operates within a MAPE-K control loop. The knowledge component maintains the architectural representation a = (M, Γ), together with system objectives and operational constraints.

*Monitoring Stage*

At each step, the system evaluates performance through a scalar objective function and computes a residual signal

$$r = J_{ref} - J_{obs} \tag{11}$$

representing deviation from expected performance. To attribute this deviation to individual modules, a sensitivity-based residual decomposition is defined. Let $\theta = (\theta_1, \dots, \theta_n)$ denote the parameter vector associated with modules $m_i \in M$. The contribution of module $m_i$ is given by

$$r_i = \left(\frac{\partial J}{\partial \theta_i}\right) \cdot \delta\theta_i \tag{12}$$

where $\delta\theta_i$ represents the deviation of the module's effective behavior from its nominal condition. When direct parameter deviations are not measurable, $\delta\theta_i$ can be approximated using local signals such as estimator innovations, actuator saturation error, or tracking residuals. This definition ensures that $r_i$ represents a first-order contribution of module $m_i$ to performance degradation and provides a consistent basis for comparison across modules. Module-level status indicators $\kappa_i$ (e.g., availability, confidence, constraint violations) are also collected.

*Analysis Stage*

The architecture is modeled as a directed graph $G = (M, \Gamma)$, with weighted adjacency matrix $W \in \mathbb{R}^{nxn}$, where

$$W_{ji} = \omega_{ji} \geq 0 \tag{13}$$

represents the influence of module $m_j$ on module $m_i$. The weights are normalized such that the spectral radius $\rho(W) < 1$, ensuring bounded propagation. Residual contributions propagate through the graph according to

$$z = (I - W^{T})^{-1} r \tag{14}$$

where $I$ denotes the identity matrix**.** $r = (r_1, \dots, r_n)^{T}$ is the vector of local residuals and z is the aggregated module influence vector.

This formulation captures multi-hop propagation of performance degradation. The Neumann-series expansion is given by

$$(I - W^{T})^{-1} = I + W^{T} + (W^{T})^2 + \dots \tag{15}$$

This Neumann-series expansion shows that $z_i$ aggregates upstream fault contributions along directed paths, providing a principled basis for multi-hop attribution. A candidate set of modules is then defined as

$$C = \{ m_i \in M \text{ such that } |z_i| \geq \tau \} \quad (16)$$

where $\tau$ is a predefined threshold. Alternatively, the top-k modules ranked by $|z_i|$ may be selected.

*Planning Stage*

The system state is defined as

$$s = (M, \Gamma, z, r, k) \quad (17)$$

Let $A(m_i)$ denote the set of admissible actions for module $m_i$, including parameter updates, module substitution, activation or deactivation, and dependency reconfiguration consistent with $\kappa_i$. For each candidate module $m_i \in C$ and action $a_k \in A(m_i)$, the expected performance improvement is approximated using a local linear model:

$$\Delta_k(m_i) \approx -\left(\frac{\partial J}{\partial \theta_i}\right) \cdot \Delta\theta_i^{\,k} \quad (18)$$

where $\Delta\theta_i^{\,k}$ represents the change induced by action $a_k$. For structural actions, this quantity may be estimated using local simulation, empirical evaluation over a short horizon, or precomputed performance tables.

*Stability-Aware Action Filtering*

To ensure safe operation, candidate actions must satisfy a stability condition. Locally approximate the system by

$$\dot{x} = f(x;\, a, \theta) \quad (19)$$

with linearization

$$\dot{x} \approx A(a, \theta)\, x \quad (20)$$

An action $a_k$ is considered admissible only if the resulting system matrix $A_k$ satisfies

$$\max Re(\lambda(A_k)) \leq -\varepsilon \quad (21)$$

for some margin $\varepsilon > 0$ where $Re(\cdot)$ denotes the real part operator and $\lambda(A_k)$ denotes the eigenvalues of $A_k$. When a model is not available, a Lyapunov-like condition is used. Let

$$V(x) = ||x - x_{ref}||^2 \quad (22)$$

where $V(x)$ is a Lyapunov candidate function measuring deviation from the reference state. Approximated as

$$\Delta V \approx V(x_{t+1}) - V(x_t) \quad (23)$$

An action is considered stability-preserving if

$$E[\Delta V\, |a_k] \leq -\alpha\, ||x_t||^2 \quad (24)$$

for some $\alpha > 0$, evaluated over a short horizon, where $\Delta V$ denotes the estimated one-step change in $V$ and $E[\cdot]$ denotes the expectation operator. The admissible action set is therefore restricted to stability-preserving actions.

*Action Selection*

Each action is assigned a score

$$Score(m_i, a_k) = z_i \cdot \Delta_k(m_i) - \lambda \cdot Cost(a_k) \quad (25)$$

where $z_i$ represents the global influence of module $m_i$, $\Delta_k(m_i)$ is the estimated improvement, $Cost(a_k)$ captures computational or switching penalties, and $\lambda > 0$ is a weighting parameter.

The selected action is $$a^* = argmax\ Score(m_i, a_k) \quad (26)$$

subject to system constraints. $$C(a, \theta, e) \leq 0 \quad (27)$$

*Switching Regularization*

To prevent oscillatory reconfiguration, a minimum dwell time is enforced. Let $t_{last}$ denote the time of the previous architectural change. A new change is allowed only if

$$t - t_{last} \geq T_{min} \quad (28)$$

Alternatively, a hysteresis condition may be applied using two thresholds $\tau_{on}$ and $\tau_{off}$ with $\tau_{on} > \tau_{off}$.

*Execution Stage*

During execution, the selected action updates parameters or modifies the architecture, after which monitoring resumes. The MAPE-K reasoning process is summarized in Algorithm 1.

**Algorithm 1: Architecture-Aware Reasoning Procedure (MAPE-K)**

---

Input: Architectural representation a = (M, Γ), weight matrix W, reference performance $J_{ref}$, module status indicators $\{k_i\}$
Output: Selected adaptation action $a^*$

---

// MONITOR: Residual Computation and Decomposition
1: Observe system performance $J_{obs}$
Compute global residual: $r \leftarrow J_{ref} - J_{obs}$ (Eq. 11)
2: For each $m_i \in M$:
Compute module-level residual contribution:
$r_i \leftarrow (\partial J/\partial \theta_i) \cdot \delta\theta_i$ (Eq. 12)
Collect module status indicator $k_i$:

// ANALYZE: Dependency-Aware influence Propagation
3: Construct weighted adjacency matrix $W \in \mathbb{R}^{n\,x\,n}$
where $W_{ji} = \omega_{ji} \geq 0,\ with\ \rho(W) < 1$ (Eq. 13)
4: Compute aggregated influence vector:
$z \leftarrow (I - W^T)^{-1} r$ (Eq. 14)

5: Form candidate set:
$C \leftarrow \{ m_i \in M \mid |z_i| \geq \tau \}$ (or select top-k modules ranked by $|z_i|$) (Eq. 16)

// PLAN: Stability-Aware Action Scoring and Selection
6: Define system state: $s \leftarrow (M, \Gamma, z, r, k)$ (Eq. 17)
7: For each $m_i \in C, for\ each\ a_k \in A(m_i)$ :
a) Estimate performance improvement:
$\triangle_k (m_i) \leftarrow -\left(\frac{\partial J}{\partial \theta_i}\right) \cdot \triangle \theta_i^k$ (Eq. 18)
b) Check stability admissibility:
If model available:
verify $max\ Re(\lambda(A_k)) \leq -\varepsilon$ (Eq. 21)
Else:
verify $E[\Delta V \mid a_k] \leq -\alpha\, ||x_t||^2$ (short horizon) (Eq. 24)
Discard $a_k$ if stability condition is not satisfied
c) Enforce switching regularization:
If $t - t_{last} < T_{min}$, skip architectural changes (Eq. 28)
d) Compute action score:
$Score(m_i, a_k) = z_i \cdot \Delta_k(m_i) - \lambda \cdot Cost(a_k)$ (Eq. 25)
8: Select optimal action:
$a^* = argmax\ Score(m_i, a_k)$ subject to $C(a, \theta, e) \leq 0$ (Eq. 26-27)

// EXECUTE: Apply Action and Resume
9: Apply a*: update $\theta_i$ or modify (M, Γ); record $t_{last} \leftarrow t$

---

### *4.3 Experimental Scenarios*

The experiment evaluates three operating conditions. Under nominal operation, low sensor noise and accurate modeling allow baseline architecture $a_1$ to perform adequately. The second condition introduces inertial measurement unit (IMU) drift, degrading orientation estimates and challenging the estimation sub-graph. The third introduces a left-wheel actuator fault causing asymmetric torque loss and challenging the control sub-graph. Table 1 summarizes the evaluated configurations using Extended Kalman Filter (EKF) estimation, Global Positioning System (GPS) fusion, and Linear Quadratic Regulator (LQR) control.

Table 1. Architectural configurations evaluated in the trajectory tracking experiment

| Config. | Estimation | Control | Fault Scenario |
|---|---|---|---|
| $a_1$ | Odometry(encoders) | LQR | Baseline/nominal |
| $a_2$ | EKF + GPS fusion | LQR | Sensor fault (IMU drift) |
| $a_3$ | EKF + GPS fusion | AsymLQR | Actuator fault (left wheel) |

Fig. 2 shows tracking performance under nominal conditions using encoder-based odometry and Linear Quadratic Regulator (LQR) control. The robot closely follows the spline-interpolated reference path, confirming that the baseline architecture is adequate when system assumptions hold.

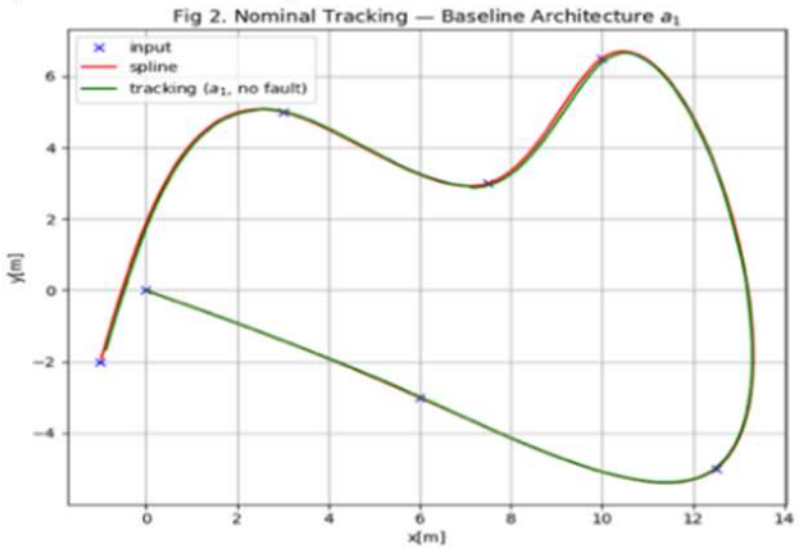

**Figure 2.** Trajectory tracking under nominal conditions. The spline reference (red) and robot trajectory (green) show accurate tracking under the baseline architecture $a_1$.

### *4.4 Adaptation Results*

Fig. 3 shows tracking performance under both fault conditions with a fixed architecture. In both cases, deviations reflect architectural limitations rather than tuning deficiencies: under IMU drift, corrupted heading measurements cause pose estimates to diverge in a manner that controller retuning cannot correct; under the actuator fault, asymmetric torque loss produces systematic path curvature that a symmetric LQR cannot structurally compensate. These failure modes motivate the architectural transitions that follow, in which the MAPE-K loop reasons over the architectural graph and selects structurally appropriate module configurations rather than attempting parameter recovery only.

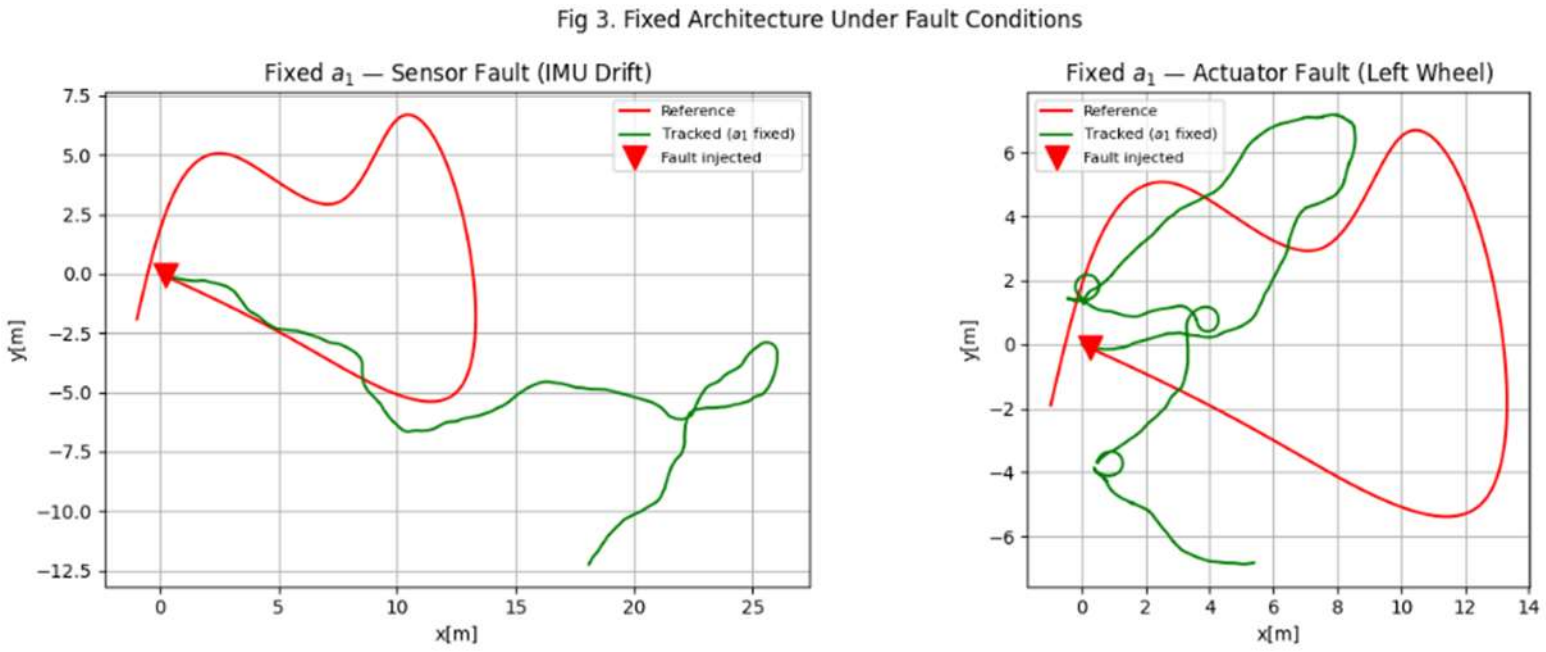

**Figure 3.** Baseline tracking under degraded conditions with fixed architecture. Sensor fault (left) and actuator fault (right) both produce growing deviations between reference (red) and tracked (green) trajectories.

Fig 4. Architecture-Aware Recovery Under Fault Conditions

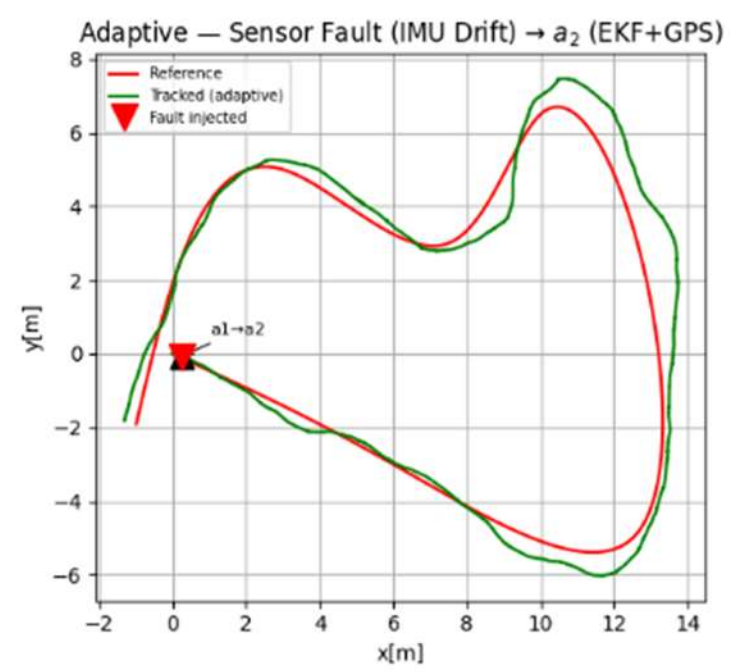

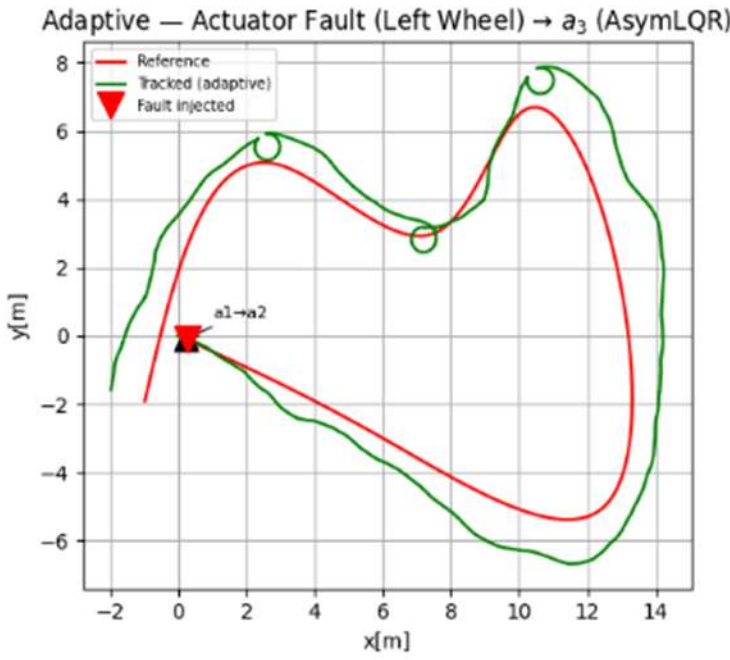

**Figure 4.** Tracking with architecture-aware adaptation enabled. Sensor fault (left) triggers transition $a_1 \rightarrow a_2$; actuator fault (right) triggers $a_1 \rightarrow a_3$. Both produce improved accuracy relative to Fig. 3

### *4.4.1 Fault-Specific Module Influence and Architecture Switch*

Fig. 5 shows normalized module influence measures $z_i$ for the estimator ($z_{estimator}$, blue) and actuator (z_act,l, green) channels over the trajectory, with adaptation threshold $\tau = 0.55$ (dashed red) and a 35-step warmup period (dotted vertical line) during which adaptation is suppressed. $\tau$ and W are design parameters set by the system architect based on knowledge of the system's dependency structure and fault characteristics; $\tau = 0.55$ was selected empirically to discriminate fault-active from nominal z-channel behavior, and W encodes known information flow relationships between modules. Sensitivity analysis and systematic identification of these parameters from operational data are directions for future work. Under the sensor fault (left), $z_{estimator}$ rises above threshold shortly after fault injection, triggering the $a_1 \rightarrow a_2$ transition. The estimator channel continues to fluctuate near threshold, consistent with persistent IMU degradation, while the actuator channel remains low. Under the actuator fault (right), $z_{act,l}$ becomes dominant, triggering the $a_1 \rightarrow a_3$ transition and activating AsymLQR. The estimator channel is elevated but secondary, confirming correct fault attribution to the actuator sub-graph. These results demonstrate that dependency-weighted influence aggregation enables fault-specific localization and targeted reconfiguration.

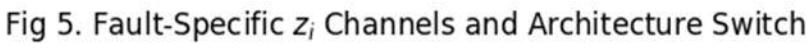

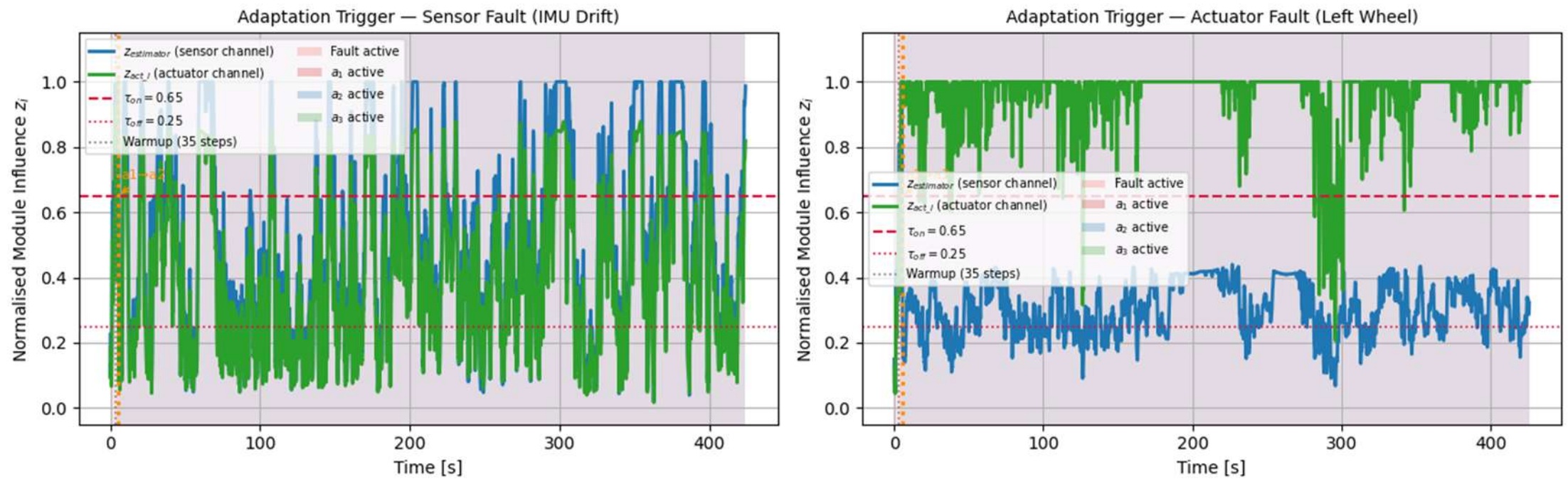

**Figure 5.** Module influence channels $z_i$ and architecture switch events (Eqs. 11–14). Sensor fault (left) shows estimator channel dominance; actuator fault (right) shows actuator channel dominance. Threshold $\tau = 0.55$ (dashed red); transitions $a_1 \rightarrow a_2$ and $a_1 \rightarrow a_3$ occur immediately after warmup.

### *4.4.2 Tracking Error: Fixed vs. Adaptive Architecture*

Fig. 6 compares Euclidean tracking error for the fixed baseline $a_1$ and the adaptive system under both faults. Under the sensor fault (left), the fixed architecture accumulates error reaching approximately 24 m by $t \approx 400$ s; the adaptive system maintains error below 1.5 m throughout following the switch to $a_2$. Under the actuator fault (right), the fixed architecture peaks near 13 m at $t \approx 130$ s and oscillates around 8 m thereafter; the adaptive system transitions to $a_3$, where AsymLQR compensates for asymmetric torque loss, maintaining error below 1.5 m with only a brief switching transient. These results confirm that architecture-aware adaptation substantially outperforms parameter-only approaches under structural fault conditions.

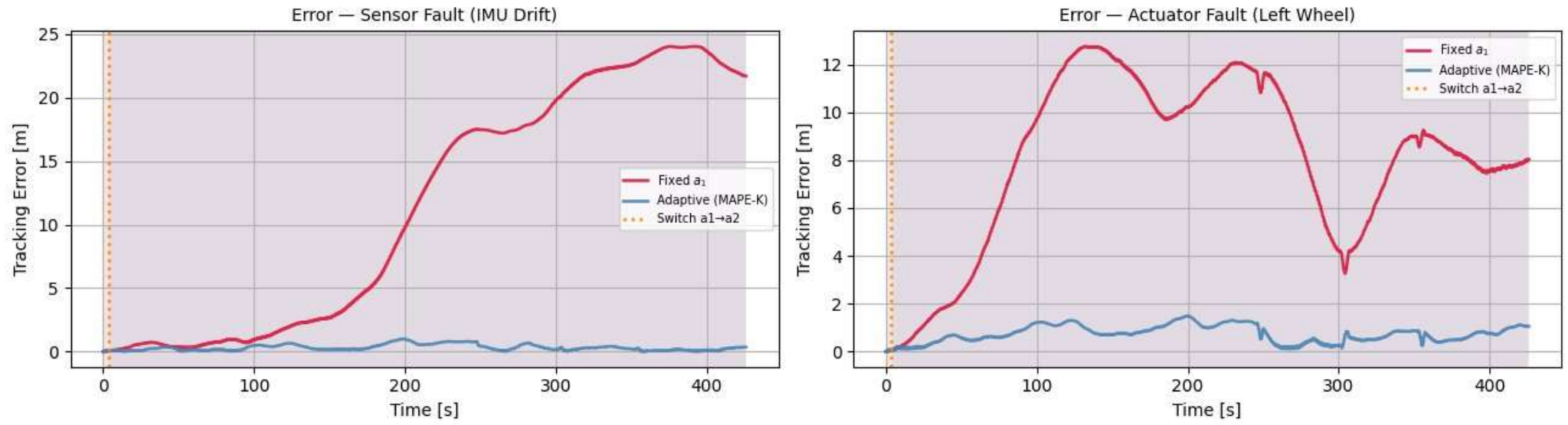

**Figure 6.** Tracking error comparison under fault injection. Left: sensor fault (IMU drift); Right: actuator fault (left wheel). Fixed architecture $a_1$ (red) produces unbounded error; the adaptive system (blue) maintains near-nominal performance after architecture switch (orange dotted line).

## 5. Conclusion

This work introduced an approach for architecture-aware autonomy in which a system reasons over an explicit, graph-based representation of its modules and dependencies. In contrast to parameter adaptation within a fixed structure, the proposed approach treats the architecture as a runtime data structure, enabling the MAPE-K loop to detect performance deviations, localize degradation via dependency-weighted influence measures, and select targeted reconfigurations within a set of candidate architectures subject to stability and operational constraints. Candidate actions are screened using a stability-aware mechanism that enforces eigenvalue-margin conditions when a model is available, and reverts to a short-horizon Lyapunov-based criterion otherwise, ensuring that transitions do not destabilize the system. In experiments on a differential drive robot subjected to IMU drift and a left-wheel actuator fault, the fixed-architecture baseline accumulated tracking errors of 24 m and 13 m, respectively, whereas the adaptive system maintained error below 1.5 m in both scenarios by transitioning to structurally appropriate configurations. A minimum-dwell-time regularization further prevented oscillatory switching during fault transients. These results show that reasoning over system structure, with stability preservation as a first-class constraint, enables a qualitatively different form of robustness that persists even when the assumptions of the original architecture fail. Future directions include handling simultaneous or cascading faults, scaling to larger module graphs, validating the approach on physical hardware, and replacing the analytical scoring function with a learned adaptation policy trained on residual-based graph representations to generalize reconfiguration decisions beyond design-time fault scenarios.

**Acknowledgements:**

The authors thank the Johns Hopkins University and the Johns Hopkins Applied Physics Laboratory for fostering a culture of innovation, particularly Jason Leggett, Justin McGrath, and Tony Johnson for their support of this work.